\documentclass[prb,aps,superscriptaddress,twocolumn,floatfix,showpacs ]{revtex4-1}

\usepackage{graphicx}
\usepackage{epstopdf}

\usepackage{color}

\usepackage[utf8]{inputenc}

\begin{document}
\title{Detailed investigation of the phase transition in K$_{x}$P$_4$W$_{8}$O$_{32}$ and 
experimental arguments for a charge density wave due to hidden nesting}

\author{Kamil Kolincio} 
\affiliation{Laboratoire CRISMAT, UMR 6508 du CNRS et de l'Ensicaen, 6 Bd Marechal
Juin, 14050 Caen, France.}
\affiliation{Faculty of Applied Physics and Mathematics, Gdansk University of Technology,
Narutowicza 11/12, 80-233 Gdansk, Poland}

\author{Olivier Pérez}
\author{Sylvie Hébert}
\affiliation{Laboratoire CRISMAT, UMR 6508 du CNRS et de l'Ensicaen, 6 Bd Marechal
Juin, 14050 Caen, France.}
\author{Pierre Fertey}
\affiliation{Société civile Synchrotron SOLEIL, L'Orme des Merisiers, Saint-Aubin - BP 48, 91192 GIF-sur-YVETTE, France.}
\author{Alain Pautrat}
\affiliation{Laboratoire CRISMAT, UMR 6508 du CNRS et de l'Ensicaen, 6 Bd Marechal
Juin, 14050 Caen, France.}

\begin{abstract}
Detailed structural and magnetotransport properties of the monophosphate tungsten bronze 
K$_{x}$(PO$_{2}$)$_{4}$(WO$_{3}$)$_{8}$ single crystals are
reported. Both galvanomagnetic and thermal properties are shown to
be consistent with a charge density wave electronic transition due to hidden nesting of quasi - 1D portion of the Fermi surface. We also observe the enhancement of electronic anisotropy due to reconstruction of the Fermi surface at the Peierls transition. The resistivity presents a thermal hysteresis  suggesting a first order
nature characteristics of a strong coupling scenario.
  However, other measurements such as the change of carriers density
demonstrate a second order Peierls scenario with weak coupling features.
We suggest that the structural transition driven by the residual strain
in the K - P - O environment is responsible for the resistivity
hysteresis and
  modifies the Fermi surface which then helps the rise to the second
order Peierls instability. 
\end{abstract}

\maketitle

\section{Introduction}
The low dimensional oxides have always attracted a wide attention due to their unconventional physics. The low dimensional electronic structure leads to the anisotropy of transport and thermoelectric properties and/or to electronic instabilities, including modulation of electronic or spin structure, known as Charge (or Spin) Density Waves. Low dimensionality is also an important ingredient for superconducting transition in copper oxides, and the interplay of CDW and superconductivity emerges as one of the central questions in that field \cite{Leroux2012}. 
K$_{x}$P$_4$W$_{8}$O$_{32}$ belongs to the monophosphate tungsten bronzes (MPTB) family described by the general formula A$_x$(PO$_2$)$_4$(WO$_3$)$_{2m}$ ($m$ being an integer, A = Na, K, Rb, Pb). The (PO$_2$)$_4$(WO$_3$)$_{2m}$ crystal structure is built by perovskite ReO$_3$ type infinite layers of corner sharing WO$_6$ octahedra connected with PO$_4$ tetrahedra \cite{roussel2001}. The regular
family members possess quasi-pentagonal tunnels large enough to accommodate additional cations which results in change of  their section to quasi – hexagonal \cite{domenges1988}. Depending on the nature of the tunnels, monophosphate tungsten bronzes are classified as MPTBp and MPTBh, respectively. K$_{x}$P$_4$W$_{8}$O$_{32}$ belongs to the MPTBp for $0 \leq x \leq 0.1$ and to MPTBh $0.7 \leq x \leq 2$ \cite{roussel19991}. It should be noted that the crystal structure of MPTB is similar to Mo$_4$O$_{11}$ Magneli phases, built of MoO$_6$ octahedra perovskite – type layers isostructural to WO$_6$ and MoO$_4$ tetrahedra playing the same role in connecting Mo – O slabs as PO$_4$ in MPTB \cite{foury20021,Canadell1989}.
The MPTB low $m$ members exhibit the 2D electronic character due to 5d conduction electrons
concentrated in WO$_6$ layers \cite{domenges1985}. The electrons are donated by PO$_4$ groups (each
group donates one electron), which play the role of charge reservoir \cite{Roussel20001}. Since
the quantity of the PO$_4$ groups is identical for each member, the number of conduction electrons
per unit cell is independent of $m$ and equal to 4 for undoped compounds. In K$_{x}$P$_4$W$_{8}$O$_{32}$  the conduction electrons density is increased 
in comparison to its parent structure, P$_4$W$_{8}$O$_{32}$  due to presence of K atoms, with valence  electron transferred to conduction band. The number of electrons per unit cell is then $4+x$.
The undoped compounds exhibit anomalies in their resistivity associated with CDW transition due to subsequent nesting of quasi 1D portions of the Fermi Surface (FS). Such behavior has been explained in the framework of hidden nesting scenario by Canadell and Whangbo \cite{canadell1991}. For the low MPTBp members, the electronic and structural transitions are observed at critical temperatures varying with $m$ \cite{roussel20002}. From X-Ray and diffuse scattering, the modulation vectors associated with the structural distortions are found to be in general agreement with the calculated nesting vector, giving support to a Peierls scenario \cite{foury20021}. On the other hand, the genuine nature of transitions in MPTBh
compounds is not clarified and remains one of the most interesting open questions in this
family.
Some of the physical properties of K$_{x}$P$_4$W$_{8}$O$_{32}$  were studied over a
decade ago.  Only a single anomaly has been observed in this compound \cite{Roussel1997}, 
what is in contrast to the
undoped $m$ = 4, where two transitions were found 
\cite{dumas2002}.
This anomaly was observed for $x$
varying from 0.86 to 1.94 with a maximum of $T^*$ = 170 K for $x$ = 1.30 \cite{Drouard1999}. 
The anomaly was also found in the thermoelectric power (TEP) at temperatures corresponding to values of $T^*$ found for transition in resistivity curves. 
Moreover, the X-ray diffuse scattering experiments proved the
existence of the long range order – structural modulation with commensurate wave vector $q =
(0.5; 0; 0)$ \cite{Dusek2002,Drouard1999}, unaffected by $x$. These results, together with the existence of single electronic gap\cite{Haffner2001} revealed via the infrared reflectivity measurements could be partially explained by a Peierls transition forecast by electronic structure calculations recalled above\cite{canadell1991}. On the other hand, the fact that the $q$ - vector does not vary with $x$, thus with the band filling was
a source of controversy about the origin of the transition\cite{Drouard1999}. Furthermore,  Bondarenko \textit {et al}~ \cite{bondarenko2004} suggested, that the relatively large anomaly observed in the specific heat at $T^*$ is  not consistent with a CDW transition but rather indicative of structural - only character of the transition. An alternative mechanism was proposed by Dusek \textit {et al}~ \cite{Dusek2002}, explaining the transition as driven by
the strain between potassium and the PO$_4$
tetrahedra causing displacement of K, P and O atoms. 
The nature of transition observed in this material is then not clear.  The scope of this work is to 
complement the information established beforehand with the new results and shed a new light on the problem of presumed CDW existence in
K$_{x}$P$_4$W$_{8}$O$_{32}$. In a broader perspective, this sample illustrates  that a weak coupling CDW can survive even if a part of the lattice is significantly distorted.
\section{Experimental}
The single crystals with the typical size of 5 mm x 1 mm x 500 $\mu$m were grown using chemical vapor transport method \cite{Giroult1982}. The potassium content was determined by EDS performed with
SEM FEI XL 30 FEG.
High resolution X-Ray diffuse scattering was studied on a selected high quality single crystal with a size of
0.19 x 0.16 x 0.015 mm$^3$. Scans were executed using monochromated radiation with
wavelength $\lambda$ = 0.50718 \r{A} and beam size of 200 $\mu$m x 200 $\mu$m. The aim of the diffraction experiment being the measurement of a maximum of satellite reflections associated to the transition, data collection at low
temperature (down to $T$ = 36 K) was accomplished with the maximum flux of the beamline but also with an attenuator for limiting possible saturation of the CCD detector by the strong main reflections. The experiment was performed both at
room temperature (RT) and at $T$ = 45 K, thus far above and deep below the transition. The samples were cooled using a He gas blower. The electrical resistivity was measured using four probe method. The experiments were performed (when possible) in two directions: in the (ab) plane ($\rho_{ab}$) and
out of plane ($\rho_{c}$). The contact configuration proposed by Hardy \textit {et al}~ \cite{Hardy1997}, adapted for
highly anisotropic systems \cite{Warmont1998, Villard19982} was used to measure $\rho_{c}$. The electrical contacts were made by welding 99,5\% Al, 0,05 \% Si,  25
$\mu m$ thick wires to the sample surface. The magnetoresistance was measured with electric current applied in the (ab) plane and magnetic field $B$ maintained perpendicularly to the direction of electric
field, in order to preserve a constant Lorenz force. The sample was rotated by $\theta$, the angle
between $B$ and c vectors. All magnetoresistance and resistivity measurements were performed using a Physical Properties Measurement System (PPMS) from Quantum Design with a 14 T magnet and equipped with horizontal rotator.
The thermopower was measured in the (ab) plane, with constant thermal gradient maintained between two chromel - constantan thermocouples soldered to the sample surface with indium. The thermocouples were also used to measure thermoeletric voltage, and the experiment was performed in the PPMS with external setup. 
The Hall coefficient was acquired by measuring transverse voltage
between two points on the sample surface in presence of  longitudinal current and magnetic
field applied perpendicularly to the (ab) plane. The magnetoresistance contribution was subtracted
from the data.
The DC magnetization was measured within a Magnetic Properties Measurement System (Squid based MPMS from Quantum Design).
Oriented single crystal was weighted and fixed with paper and scotch in
a polyethylene straw. Constant magnetic field was applied along (ab) plane. High value of $B$ =
1 T was used in order to achieve the finest scan resolution and extract the weak sample signal from substantially stronger background.
The $I(V)$ curve was obtained by a four probe method. The current was applied along the sample
(in the (ab) plane) using Keithley 228A source, and voltage was measured by microvoltmeter. The I contacts were enhanced by welding additional AlSi wires joined together with a small drop of DuPont 6838 silver paste. The sample was kept in liquid nitrogen to ensure effective contacts cooling.

\section{Results and discussion}
\subsection{Structural analysis}
The purpose of the crystallographic study is to verify the results previously obtained by Dusek \textit {et al}\cite{Dusek2002}. In this previous study, the 
data was also achieved using synchrotron radiation, almost mandatory to solve the structure with such weak satellite peaks.  789 main and 409 satellite reflections with I $\geq 2.5 \; \sigma (I)$
were then collected at 110 K, thus close to the transition temperature.

\begin{figure}[h]
  \includegraphics[scale=0.25]{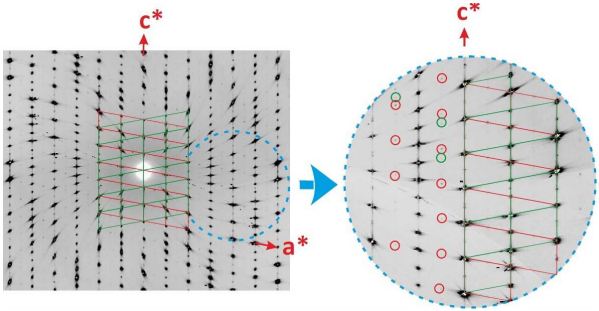}
 \caption{\label{figu-diff} (h0l)* plane at 45 K. Drawn green and red unit cells are related to the reticular pseudo-merohedral twinning but refer to the 
RT cell {\textit i.e.} to a $\approx $ 6.68 \AA, b $\approx $ 5.32 \AA, c $\approx $ 8.90 \AA\ and $\beta$ $\approx $ 100.6$^{\circ}$ . In the
zoom view, the satellite reflections leading to the doubling of the a parameter and associated to the transition are enlightened by red and green circles showing the satellites of red and green twin domains respectively.   }
 \end{figure}

The reconstructed
(h0l)$^*$ planes using the experimental frames collected with our single crystal at RT and 45 K (see figure~\ref{figu-diff}) show unexpected
extinction rules: rows running along c$^*$ with a periodicity of $\approx$ 8.9 \AA\  alternate with rows
exhibiting a periodicity of $\approx$ 17.8 \AA. This phenomenon can be interpreted using the hypothesis
of occurrence of twin domains related by a two – fold axis parallel to c; our sample exhibits then a reticular pseudo-merohedral twinning. The cell at RT was
found to be monoclinic with the following  cell parameters: a = 6.676(1) \AA, b
= 5.322(1) \AA, c = 8.899(1) \AA\  and $\beta$ = 100.637(2)$^{\circ}$. The hypothesis of a reticular pseudo-merohedral twinning has been successfully tested by a full structural refinement procedure for the RT data set. 

The experiment performed at reduced temperature $T$ = 45 K revealed the existence of weak
superlattice reflections. In fact, two sets of additional peaks were found; they are corresponding to satellites associated to each
twin domain as shown in figure~\ref{figu-diff} and they induce the doubling of the a parameter. The observation of this two sets
is a supplementary proof validating our twin hypothesis. The positions of the satellites are in agreement both with the modulation vector previously
reported by Dusek \textit {et al} ($q$ = (0.5; 0; 0)) and with calculations performed by Canadell and Whangbo \cite{canadell1991}, who
predicted this value as associated with hidden nesting of the 1D part of the Fermi surface. The diffraction pattern lead then to the following cell parameters : 
a = 13.2383(6) \AA, b = 5.2823(1) \AA, c = 8.8511(2) \AA\ and $\beta$ = 100.677(3)\AA.

The data recorded at 45 K at the Cristal beamline at Soleil contain 4113 observed reflections with  I $\geq 3 \; \sigma (I)$, 1703 reflections with $h=2n+1$ (\textit{i.e.} satellites) and 2410 reflections with
$h = 2n$ (\textit{i.e.} main); indices are referring to the unit cell above. The data set is then globally 3.4 times larger than the one collected by Dusek \textit {et al} and we are 
observing 4.15 times more satellite reflections directly related to the transition. Note that the data collection was performed deep below the transition temperature to achieve the largest possible enhancement of the satellite reflections.
 However, let us notice that our sample
exhibits a reticular pseudo-merohedral twinning; the two domains are related by a mirror perpendicular to a. 
The main reflections of both domains are fully or partly overlapped and the integration process requires additional  corrections; the intensity of the main reflections is then expected to be less reliable than the fully separated satellites 
(see figure~\ref{figu-diff}).

 The refinement procedure was achieved using Jana2006~\cite{jana2006} in the 3-dimensional supercell with the space group $P2_1$ \ following the analysis 
of Dusek \textit {et al}. The structure is fully described by 4 W, 2 P, 2 K and 16 O atoms. The atomic displacement parameters (ADP) were considered
isotropic for O atoms and anisotropic in the other cases. Atoms related by inversion center in the centrosymmetic $P2_1/m$ approximation are restricted
to have the same ADP. The occupancy of the two K atoms is refined; they exhibit a partial occupancy of 0.291(4) and 0.279(4) leading to the chemical composition
K$_{1.14(2)}$P$_4$W$_{8}$O$_{32}$, \ close to the expected formula. The final agreement factors, with 109 refined parameters, are 6.97\% for the 4113 reflections, 6.78\% for the 2410 reflections with $h = 2n$ and 8.05\% for the 1703 reflections with $h = 2n+1$. Atomic parameters of this refinement are given in table~\ref{param} and atomic distances in table~\ref{dist}. 

\begin{figure}[h]
  \includegraphics[scale=0.35]{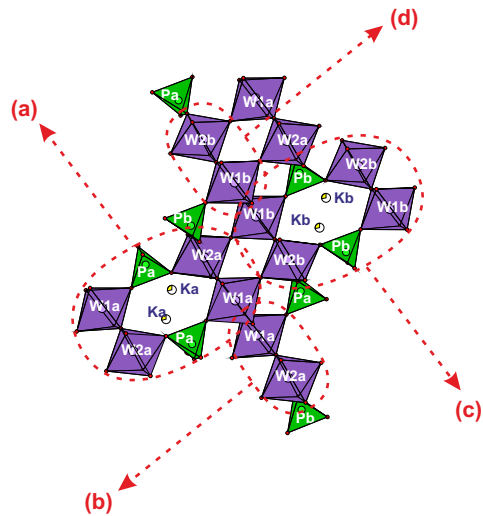}
 \caption{\label{figu-proj-b} Projection along {\textbf b} of an area of the structure exhibiting both the hexagonal tunnels and the $(WO_3)_m$ \ rows. $WO_6$
\/ and $PO_4$ \/ environments are drawn using purple and green colors respectively. The (a), (b), (c) and (d) circled zones are referring to the figure~\ref{figu-detail}.}
 \end{figure}

\begin{figure}[h]
  \includegraphics[scale=0.2]{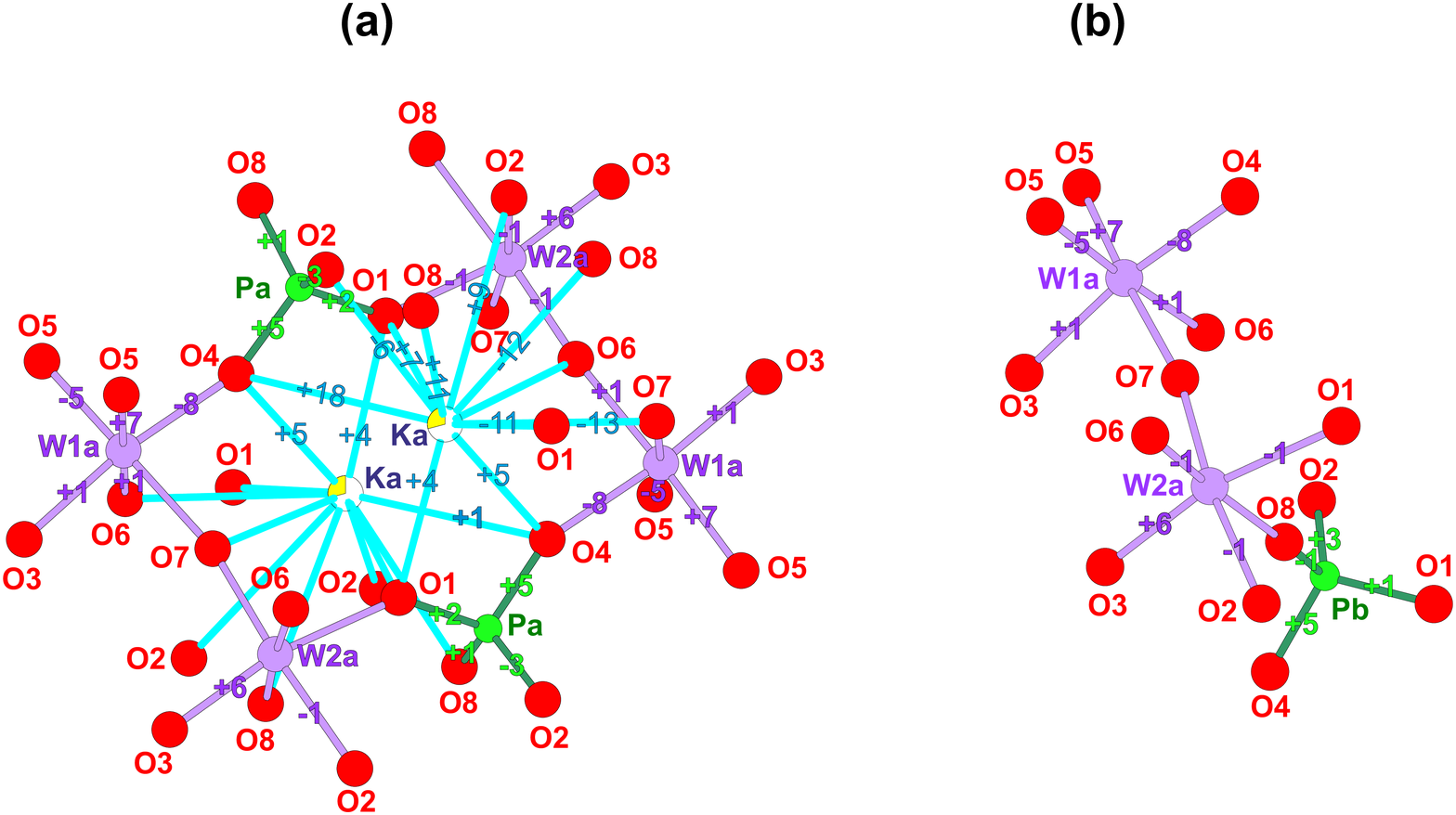} \\[0.5cm]
  \includegraphics[scale=0.2]{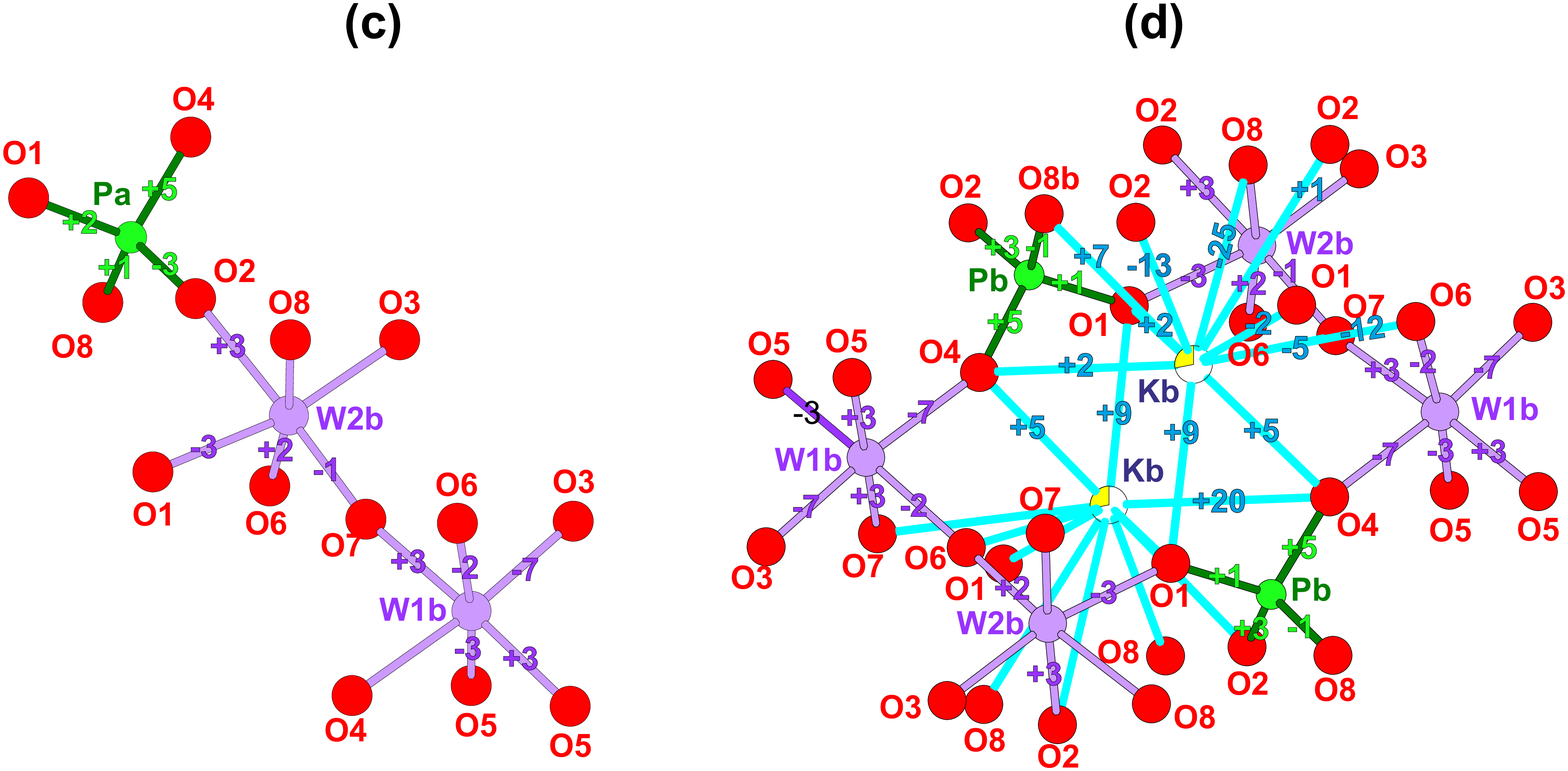} 	
 \caption{\label{figu-detail} Details of the projection drawn in figure~\ref{figu-proj-b}. Numbers reported on the boundaries are related to the deviation
observed between the room temperature and the 45 K structures for the corresponding distance; they are stated in hundredth of \AA; - sign evidences a contraction at low temperature and + sign to a dilatation. }
 \end{figure}

The structural results obtained at 45 K are compared with the analysis performed at room temperature (RT).  The deviations between the atomic positions at 45 K and RT are reported in the table~\ref{param}. The maximum atomic displacements observed
for the O, K, P and  W atoms are $\approx$ 0.19, 0.054, 0.111 and 0.034 \AA \/ respectively (see table~\ref{param}). These moderate values evidenced for W were
already reported by Dusek \textit {et al}~\cite{Dusek2002}; they did not induce any significant change in the W-W boundaries. The figure~\ref{figu-detail}
shows the main modifications observed at 45 K for the cation-oxygen distances corresponding to the structure zones indicated in figure \ref{figu-proj-b}. The superstructure observed at 45 K does not induce the large distortions in the
WO or PO bounding scheme (see mainly figure~\ref{figu-detail} b and c). The strongest effect is related to the geometry of the hexagonal windows 
(see figure~\ref{figu-detail} a and d). A dilatation of the cavities is observed in the direction of the Pa-Pa segment and a contraction in the perpendicular
direction.

\begin{table}[h]
\caption{\label{param}Positional coordinates and atomic displacement parameters}
{\tiny \begin{ruledtabular}\begin{tabular}{ccccccc} 
atom & occu& x & y& z & U$_{eq}$ \ (\AA$^2$) & $\left|\Delta (RT-45K)\right|$  \\
& pation & & & & &  (\AA) \\
\hline

  W1a  & 1         &-0.06936(3)&-0.25383(9)& 0.58862(5)& 0.00054(6)      & 0.021  \\           
  W1b  & 1         &-0.43044(3)& 0.25641(10)&-0.58770(5)& 0.00054(6)     & 0.034  \\           
  W2a  & 1         &-0.21793(4)& 0.24434(9)& 0.74920(5)& 0.00052(6)      & 0.030   \\          
  W2b  & 1         &-0.28222(3)&-0.24711(8)&-0.74907(5)& 0.00052(6)      & 0.016   \\          
  Pb   & 1         & 0.3545(2)& 0.2398(7)& 0.1290(4)& 0.0015(3)          & 0.054   \\          
  Pa   & 1         &-0.8546(2)&-0.2434(7)&-0.1298(4)& 0.0015(3)          & 0.036   \\          
  O1a  & 1         & 0.3664(6)& 0.2387(19)&-0.0386(8)& 0.0024(5)         & 0.066   \\          
  O1b  & 1         &-0.8691(6)&-0.2532(18)& 0.0379(9)& 0.0024(5)         & 0.022   \\          
  O2a  & 1         & 0.3055(6)&-0.0143(15)& 0.1582(10)& 0.0025(5)        & 0.179    \\         
  O2b  & 1         &-0.8092(7)&-0.0002(15)&-0.1660(10)& 0.0025(5)        & 0.155    \\         
  O3a  & 1         &-0.3225(6)& 0.2358(16)& 0.5760(9)& 0.0033(6)         & 0.115    \\         
  O3b  & 1         &-0.1858(6)&-0.2418(16)&-0.5767(11)& 0.0033(6)        & 0.049    \\         
  O4a  & 1         & 0.4625(6)& 0.266(2)& 0.2222(9)& 0.0030(5)           & 0.131    \\         
  O4b  & 1         &-0.9631(6)&-0.256(2)&-0.2218(10)& 0.0030(5)          & 0.010    \\         
  O5a  & 1         & 0.0112(6)&-0.0176(12)& 0.4944(9)& 0.0007(5)         & 0.190    \\         
  O5b  & 1         &-0.4997(5)& 0.0081(12)&-0.5041(8)& 0.0007(5)         & 0.057    \\         
  O6a  & 1         &-0.1439(6)& 0.4945(14)& 0.6855(9)& 0.0015(5)         & 0.046    \\         
  O6b  & 1         &-0.3518(6)&-0.5164(14)&-0.6898(9)& 0.0015(5)         & 0.145    \\         
  O7a  & 1         &-0.1364(7)& 0.0030(15)& 0.6903(10)& 0.0016(5)        & 0.057    \\         
  O7b  & 1         &-0.3651(7)&-0.0200(15)&-0.6833(10)& 0.0016(5)        & 0.129    \\         
  O8a  & 1         & 0.2976(6)& 0.4692(15)& 0.1667(10)& 0.0030(5)        & 0.104    \\         
  O8b  & 1         &-0.7880(6)&-0.4643(15)&-0.1587(9)& 0.0030(5)         & 0.182    \\         
  Ka   & 0.29(1) &-1.0697(7)&-0.2354(18)&-0.0011(10)& 0.0042(10)       & 0.111   \\          
  Kb   & 0.28(1) & 0.5706(7)& 0.246(2)& 0.0027(11)& 0.0042(10)         & 0.099   \\

\end{tabular}\end{ruledtabular}}
 
\end{table}

\begin{table}[h]
\caption{\label{dist}Interatomic distances. Symmetry codes: (i) x,y,z+1; (ii) x+1,y,z+1; (iii) -x,y-1/2,-z+1 ; (iv) x,y-1,z; (v) x,y,z-1; (vi) x-1,y,z-1; 
(vii) -x-1,y+1/2,-z-1; (viii) x,y+1,z ; (ix) -x-1,y+1/2,-z+1; (x) -x,y+1/2,-z+1; (xi) -x,y-1/2,-z-1; (xii) -x-1,y-1/2,-z-1; 
(xiii) x+1,y,z; (xiv) -x-1,y-1/2,-z; (xv) -x-1,y+1/2,-z; (xvi) -x,y-1/2,-z; (xvii) x,y-1,z-1; (xviii) -x+1,y-1/2,-z;
(xix) -x+1,y+1/2,-z; (xx) -x,y+1/2,-z; (xxi) x+1,y+1,z+1} 
{\tiny \begin{ruledtabular}\begin{tabular}{cccccc}

boundary &  d (\AA) & boundary & d (\AA) & boundary & d (\AA) \\
\hline

W1a-O3b$^{i}$ &1.920(8)	  & 	W1b-O3a$^{v}$ &1.840(7)    	     &  W2a-O1b$^{ix}$ &2.018(7) 	 \\              
W1a-O4b$^{ii}$ &1.979(8)      &         W1b-O4a$^{vi}$ &1.990(7)     &  W2a-O2a$^{x}$ &1.999(9)      \\                  
W1a-O5a&1.929(8)          &         W1b-O5b&1.833(7)         &  W2a-O3a&1.866(7)          \\                             
W1a-O5a$^{iii}$ &1.812(7)     &         W1b-O5b$^{vii}$ &1.887(7)    &  W2a-O6a&1.797(8)          \\                     
W1a-O6a$^{iv}$ &1.947(8)      &         W1b-O6b$^{viii}$ &1.921(8)   &  W2a-O7a&1.808(9)          \\                     
W1a-O7a&1.932(9)          &         W1b-O7b&1.967(9)         &  W2a-O8a$^{iii}$ &2.018(8)     \\                         
\hline                                                                                                                   
W2b-O1a$^{xi}$ &2.000(7)  	  & 	Pa-O1b&1.532(9)		     &   Pb-O1a&1.521(8)           \\                    
W2b-O2b$^{xii}$ &2.038(9)     &         Pa-O2b&1.478(9)          & Pb-O2a&1.533(9)            \\                         
W2b-O3b&1.799(8)          &         Pa-O4b&1.517(8)          &  Pb-O4a&1.520(8)            \\                            
W2b-O6b&1.824(8)          &         Pa-O8b&1.513(9)          &  Pb-O8a&1.496(9)            \\                            
W2b-O7b&1.794(9)          &                                  &                   & \        & \\                               
W2b-O8b$^{vii}$ &2.011(8)     &                                  &                         & \ & \\                           
\hline                                                                                                                   
Ka-O1b$^{xiii}$ &2.616(12) 	  & 	Kb-O1a&2.660(12)             &     & \                \\                           
Ka-O1b$^{xiv}$ &2.854(13)     &         Kb-O1a$^{xviii}$ &2.807(14)  &                          & \ \\                       
Ka-O1b$^{xv}$ &2.675(13)      &         Kb-O1a$^{xix}$ &2.734(14)    &                          \ & \ \\                       
Ka-O2a$^{xvi}$ &3.499(12)     &         Kb-O2a$^{xix}$ &2.681(14)    &                          & \ \\                       
Ka-O2b$^{xiv}$ &2.755(13)     &         Kb-O2b$^{xx}$ &3.489(12)     &                          & \ \\                       
Ka-O4b$^{xiii}$ &2.615(14)    &         Kb-O4a&2.619(14)         &                          & \ \\                           
Ka-O4b$^{xiv}$ &3.366(14)     &         Kb-O4a$^{xviii}$ &3.201(14)  &                          & \ \\                       
Ka-O4b$^{xv}$ &3.190(13)      &         Kb-O4a$^{xix}$ &3.375(14)    &                         & \ \\                        
Ka-O6a$^{xvii}$ &3.112(12)    &         Kb-O6b$^{?}$ &3.000(12)      &                         & \ \\                        
Ka-O7a$^{v}$ &2.989(12)       &      Kb-O7b$^{ii}$ &3.089(12)        &                       & \ \\                          
Ka-O8a$^{xvi}$ &3.284(12)     &      Kb-O8a$^{xviii}$ &2.899(14)     &                       & \ \\                          
Ka-O8b$^{xv}$ &2.931(13)      &      Kb-O8b$^{xx}$ &3.154(12)        &                        & \ \\                         

\end{tabular}\end{ruledtabular}}

\end{table}

The paper written by Dusek {\textit et al}~\cite{Dusek2002} provides a comparison between the average structure and the superstructure both calculated from the data collected at 110 K. Our present description details the differences observed between the structure of the fundamental state at room temperature and the structure at 45 K deep below the transition. The large number of observed independent satellite reflections (4.5 times more than in the previous study~\cite{Dusek2002}) leads
to a high degree of confidence in our result and very low standard deviations. This last point is very important since the differences observed above and below 
the transition are very weak in term of atomic positions (see table~\ref{param}) and even slighter in term of distances (see figure~\ref{figu-detail}). The obtained
results clearly validate the conclusions drawn by Dusek {\textit et al}~\cite{Dusek2002}: the main changes at 45 K are occurring for the phosphorus and potassium environment as evidence by the variations of the K-O and P-O distances. However for us, such effect does not contradict the possibility of CDW formation since weak but significant deviations ($\approx$ 0.01 to 0.02 \AA) are also observed for W-W and W-O distances.

\subsection{Physical properties}

\begin{figure}[h]
  \includegraphics[scale=0.32]{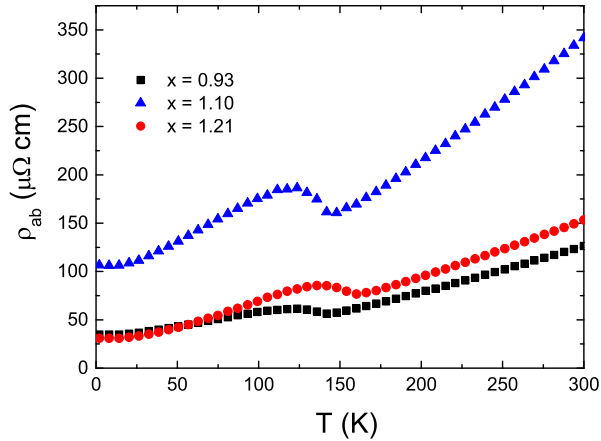}
 \caption{\label{rhoab} In - plane resistivity vs. temperature of K$_{x}$P$_4$W$_{8}$O$_{32}$ for various values of $x$}
 \end{figure}

\begin{figure}[h]
\includegraphics[scale=0.32]{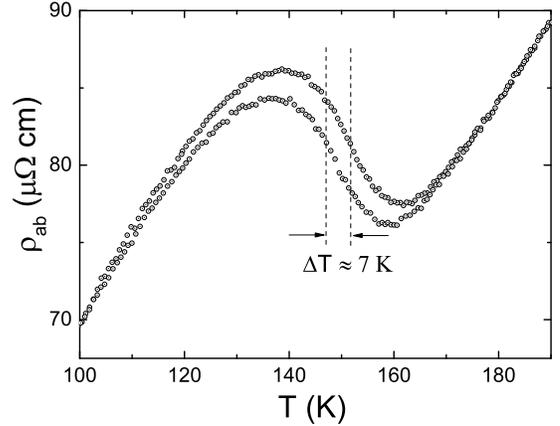}
  \caption{\label{hist} Resistivity as function of temperature around the suspected CDW transition in K$_{1.21}$P$_4$W$_{8}$O$_{32}$. The rate of temperature change during the heating and cooling procedure is 0.05 K/min.   }
 \end{figure}

The typical in - plane resistivity vs. temperature curves, measured for different $x$, are shown in fig. \ref{rhoab}. 
  As previously reported\cite{dumas2002,Drouard2000}, a clear
anomaly is observed in the 100 - 150 K range at a temperature $T^{*}$, which
depends on the potassium value. Additionally, a small thermal hysteresis is observed at the transition.  To ensure, if this effect is truly intrinsic to the sample and is not an artifact caused by insufficient thermalisation, we have repeated the scan with heating/cooling rate as low as 0.05 K/min. As a result, we have found the reproducible hysteresis, with a very similar width of $\Delta T \approx 7 $ K for $x$ = 1.21 (shown in fig. \ref{hist}) and $x$ = 1.10 and, a smaller one  ($\Delta T \approx 3 $ K) for $x$ = 0.93. 
 
\begin{figure}[h]
 \includegraphics[scale=0.32]{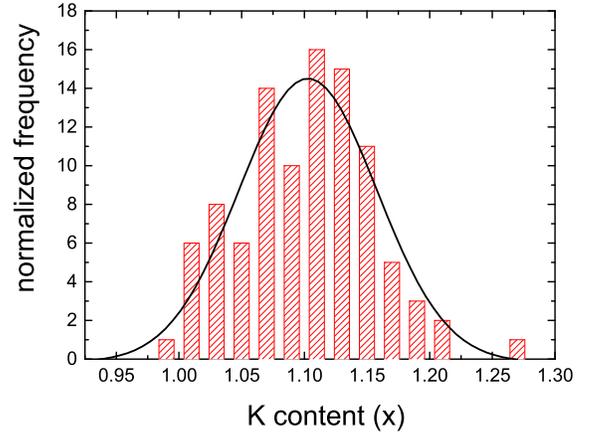}
  \caption{\label{concentration} Distribution of $x$ with K$_{x}$P$_4$W$_{8}$O$_{32}$. single crystal, and Gaussian decay fit (solid line).}
  \end{figure}

To go deeper insight into this anomaly, we have performed additional sample characterizations and analyzed in details the distribution of potassium content in the crystals. The idea behind is that any departure from
 a very precise stoichiometry could result in a broadening of the transition,
 as it is well known in high T$_c$ cuprates for instance\cite{Shi1989}. 
 The performed EDS scans reveal, that the samples composition is not perfectly homogeneous  at the sample scale. The K distribution for the $x$ = 1.10 crystal is shown in fig. \ref{concentration}. We find, that it can be described by the Gaussian decay with $x=1.10$ and FWHM = 0.11.
  Considering the $T^*(x)$ dependence \cite{Drouard2000}, one can expect a broadening of  $\Delta T^*  \approx 10 K$ corresponding to the obtained distribution width.
  Note that strictly, that although $\Delta T^*$ stands in agreement with the hysteresis size
only a broadening is expected by such a $\Delta T^*$ effect. One could consider the scenario of the current percolation in the preferential zones, which undergo the transition at various temperatures  as an explanation of the macroscopic hysteresis, however this picture would lead to the deviation from the Ohmic behavior, which is not observed in our samples. Therefore, we find this mechanism irrelevant to explain the hysteretic manner of resistivity.
Note, that the existence of hysteresis can be a fingerprint of a first-order transition, \textit{a priori}~ unexpected  in the framework of  Peierls-Frohlich theory which is a continuous second order transition \cite{Gruner1988}. The Ginzburg Landau approach predicts the first order character  of a lock - in transition separating an incommensurate state at high temperature and a commensurate state \cite{McMillan1975}. Such effects observed in various CDW materials as TaSe$_2$\cite{Moncton1977} or K$_{0.3}$MoO$_3$\cite{Fleming1985}. We find this case unlikely here, since neither transport or structural properties reveal any sign of an incommensurate CDW at higher temperatures. However it shall be noted that the first order transitions towards CDW has been reported in Lu$_5$Rh$_4$Si$_{10}$ \cite{Lue2002}, Lu$_5$Ir$_4$Si$_{10}$ \cite{Becker1999, Leroux20121, Jung2003} without any signatures of CDW phase at higher temperatures.

\begin{figure}[h]
 \includegraphics[scale=0.32]{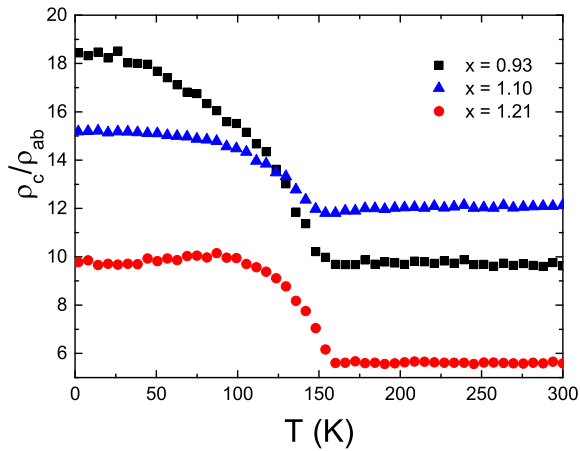}
 \caption{\label{anizotrop} Temperature dependence of $\rho_c$  / $\rho_{ab}$ ratio for various concentrations of potassium in K$_{x}$P$_4$W$_{8}$O$_{32}$}
\end{figure}

   The anisotropy in electronic properties, expected in a quasi 2D system
is inferred by measurements of the out of plane resistivity $\rho_{c}$ (fig. \ref{anizotrop}). 
This latter
shows a typical metallic behavior consistent with a coherent scattering
perpendicular to the most conducting planes and an anomaly at $T^*$.
The anisotropy of resistivity  $\rho_{c}/\rho_{ab}$ varies from 6 to 12 at
room temperature. $\rho_{c}/\rho_{ab}$ is strictly constant down to $T^*$, but remarkably grows from $T^*$ to low temperature.
This indicates a strong increase of the transport anisotropy. The reinforcement
of anisotropy has been noted in materials with CDW as NbSe$_{2}$\cite{Leblanc2010}, albeit without quantitative analysis.  The constant electronic anisotropy for $T>T^*$ is consistent with a simple metallic state. For $T<T^*$, the strong change indicates that this anisotropy changes likely due to a Fermi surface reconstruction. It shall also be noticed, that the temperature variation
of $\rho_{c}/\rho_{ab}$ is very similar to the thermal dependence of a Peierls gap. From this observation, at least we can propose that its main origin is not from the change of electronic mobility of carriers, but rather from an anisotropic change of carriers density at $T^*$. 

\begin{figure}[h]
 \includegraphics[scale=0.32]{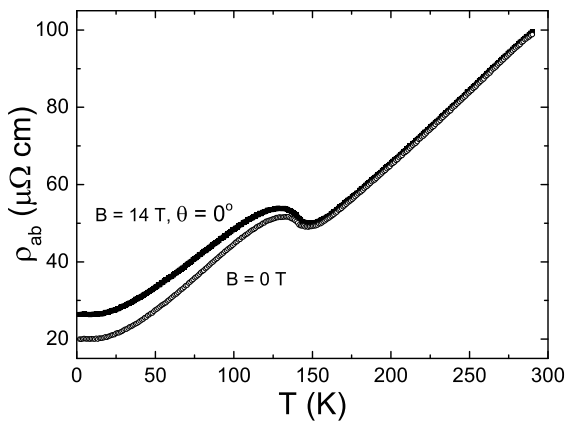}
\caption{\label{magnres1} Electrical resistivity and magnetoresistance in K$_{1.20}$P$_4$W$_{8}$O$_{32}$ as a function of temperature. $\theta$ is the angle between $B$ and c*}
 \end{figure}

Magnetotransport measurements show a positive, metallic-like, magnetoresistance (MR), which increases significantly below $T^*$, as shown in fig. \ref{magnres1}.

\begin{figure}[h]
 \includegraphics[scale=0.32]{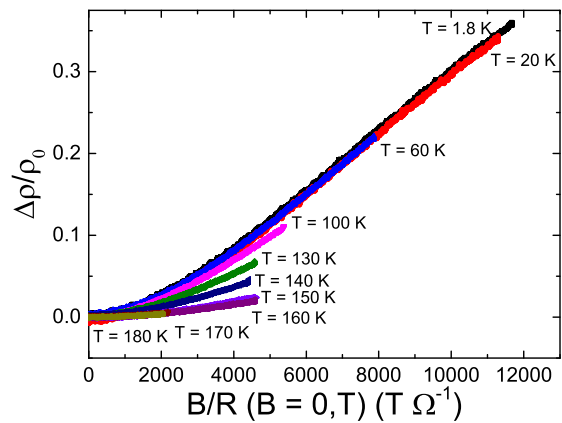}
  \caption{\label{kohler} Kohler's plot for K$_{1.20}$P$_4$W$_{8}$O$_{32}$}
 \end{figure}

The MR follows a cosine law, reaching maximum for $B\perp c$, and is almost zero when $B$ lies in the (ab) plane. The $cos \theta$ regime followed by the magnetoresistance is a typical signature of cylindrical, quasi 2D topology of the Fermi surface \cite{Pippard1989}. The MR can be discussed in the framework of the Kohler's rule, predicting that all the Kohler's plots:

\begin{equation}
\frac{\Delta\rho}{\rho_0}=F(\omega_c\tau)=F\left(\frac{H}{\rho(B=0,T)}\right)
\end{equation} 
(where $\omega_c$ is the cyclotron frequency and $\tau$ is the scattering time) follow the same curve, if $\tau$ is identical for all
carriers, and constant in all parts of Fermi surface \cite{McKenzie1998}. In fig \ref{kohler}, we show the Kohler plot for the $x$ = 1.20 sample.

One can notice that all the curves for temperatures above $T^* \approx $ 158
K fall into one line. When decreasing further the temperature, the Kohlers's rule is violated down to $T\approx$ 60 K, where it is again satisfied. Different causes can lead to a breakdown of Kohler's rule\cite{McKenzie1998}. One possibility is the formation of a density wave at a critical temperature $T_c$ and of a concomitant change of electronic structure, as observed in other materials with Peierls instability such as NbSe$_3$ \cite{Yasuzuka2005}. In this case, the change is expected to be significant down to roughly $T_c/2$ where the gap starts to be fully open \cite{Gruner1988}. This is consistent with our observations.

\begin{figure}[h]
  \includegraphics[scale=0.3]{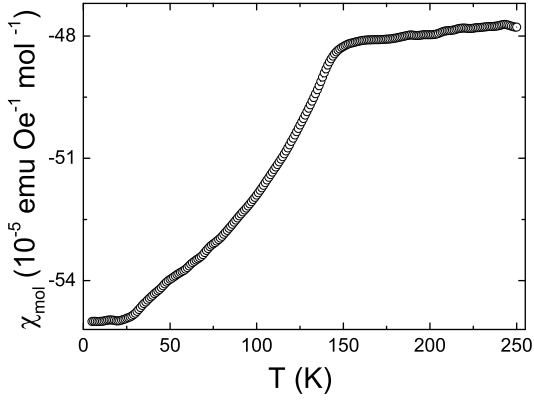}
 \caption{\label{chimol} Temperature dependence of the in plane molar susceptibility of K$_{1.16}$P$_4$W$_{8}$O$_{32}$ corrected from background signal and the Curie - Weiss component.}
 \end{figure}

\begin{figure}[h]
\includegraphics[scale=0.3]{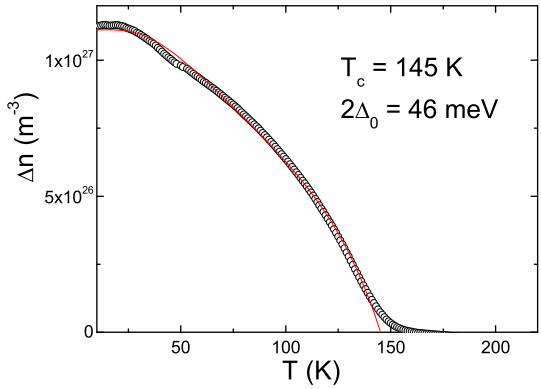} 
 \caption{\label{bcs} Change of estimated carriers density as function of the temperature (dots). The solid line is a fit with a BCS-like expression (see text for details). }
 \end{figure}

 The magnetic properties of a large oriented single crystal K$_{1.16}$P$_4$W$_{8}$O$_{32}$ with mass $m$ = 0.0744 g have been measured. The background magnetization due to the sample holder was measured and carefully removed. Then the parasitic Curie - Weiss contribution ($\chi_C$) due to paramagnetic impurities was quantified and subtracted from the data. The obtained magnetic susceptibility is shown in the figure \ref{chimol}.  Above $T^*$, $\chi$ is almost temperature independent, while for $T<T^*$ one observes a notable drop of magnetic susceptibility, reminiscent to the decrease of $\chi$ due to the gap opening and CDW condensation of free electronic carriers reported in known CDW materials as Tl$_{0.3}$MoO$_3$  \cite{collins1985}, $\gamma-$Mo$_4$O$_{11}$  \cite{Schlenker1986}, or in undoped MPTBp $m$ = 4  \cite{Teweldemedhin1992}. 
To analyse the data, one has to consider the following contributions in a metallic-like
sample:
\begin{equation}
\chi = \chi_{ion}+\chi_{V}+\chi_{C}+\chi_{P}+\chi_{L},
\end{equation}
The $\chi_{ion}$ -  the ion core diamagnetism, and $\chi_V$ – Van Vleck paramagnetism of ions, were assumed to be constant with temperature. 
From the remaining ingredients: $\chi_P$, the Pauli paramagnetism of conduction electrons, and Landau diamagnetism of carriers ($\chi_L$) we have deduced the number of condensed electrons as a function of temperature, shown in fig. \ref{bcs}. The maximum $\Delta n \approx 1.12 \cdot 10^{27}m^{-3}$ corresponds to 15\% of the electronic density calculated from chemical formula ($n_0$ = 8.16 $\cdot 10^{27}m^{-3}$).
In a BCS - Peierls description of the CDW transition, the number of condensed electrons follows the temperature variation\cite{Khasanov2008}:
\begin{equation}
n_{condensed}=n_{0}\frac{\Delta(T)}{\Delta_{0}}tanh \left(\frac{\Delta(T)}{2k_{B}T}\right) \label{eq:nbcs}
\end{equation} 
with the electronic gap:
\begin{equation}
\Delta(T)= \Delta_{0}tanh \left( \frac{\pi k_{B}T_{C}}{\Delta_{0}}\sqrt{1.018\left( \frac{T_{C}}{T}-1 \right)} \right)\label{eq:delta}
\end{equation} 
where $\Delta_0$ is the electronic gap at $T$ = 0 K.
Here, we find a good fit with a
BCS-mean weak coupling expression. A slightly better fit can be proposed when removing the low
temperature part ($T<40$ K) where the raw data is affected by the
parasitic Curie Weiss contribution.  However, since this does not change
significantly the fitting parameters, we prefer to show the fit on all
the $T<150$ K part. We have found the electronic gap 2$\Delta_0$ = 46 meV for $x$ = 1.15 , which converges with the value predicted
by mean – free model in limit of the the weak electron – phonon coupling 2$\Delta=$ 3.52 $kT^*\approx$ 44.6 meV. Here, we have to mention the work of Haffner \textit {et al} \cite{Haffner2001} who have measured the infrared absorption of K$_{1.30}$P$_4$W$_{8}$O$_{32}$. They report at low temperature absorption peaks which were tentatively attributed  to single-particle excitations across a gap. In this framework, a value $2\Delta\approx$ 43 meV can be deduced. It is worth noting that this value compares extremely well with our magnetization data and this indicates that condensed carriers are effectively at the origin of these absorption peaks as proposed by Haffner \textit {et al}.

\begin{figure}[h]
 \includegraphics[scale=0.3]{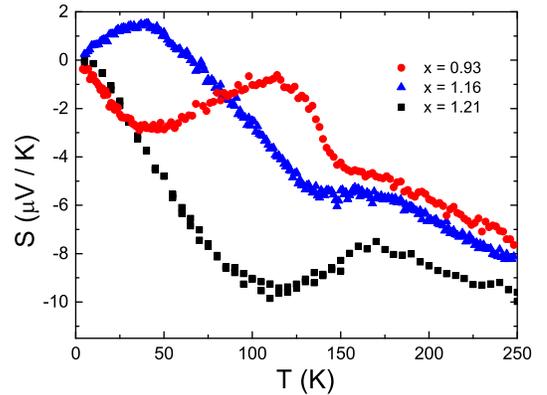}
  \caption{\label{seebeck} Normalized thermopower as a function of temperature in K$_{x}$P$_4$W$_{8}$O$_{32}$ for different potassium concentrations}
 \end{figure}

To go deeper insight into the change of carrier density upon the transition, we have conducted the study of thermopower with various potassium concentrations. The temperature dependence of Seebeck coefficient for different values of $x$ is shown in fig. \ref{seebeck}.
 All $S(T)$ present a transition between two metallic regimes, with linear
$S(T)$ curves, at low and high temperatures. In the transition region, depending on
$x$, the transition is associated to an increase of $S$ ($x$ = 0.93), a
decrease of $S$ ($x$ = 1.21), or for the intermediate doping ($x$ = 1.16), a
constant value of $S$. The thermopower even turns positive at low temperature for $x$ = 1.16. The above described behaviors are in good agreement with the results reported previously \cite{Drouard2000,Roussel1997}. The maximum in $S(T)$ close to $S$ = 0, indicates enhancement of the hole influence for $x$ = 0.93. On the other hand, electrons remain dominant carriers in $x$ = 1.21 compound. Roussel \textit {et al} \cite{Roussel1997}  suggested, that strong
change of slope and existence of an extremum, either in positive or negative direction might result from Peierls transition
with high – mobility carriers left in the pockets remaining after imperfect nesting of the Fermi surface. To follow the thermal evolution of the effective electronic density $n_{eff}$ containing both the electrons and holes contributions (with opposite signs), we have applied the semi - classical 3D model of quasi - free electrons \cite{Palecchi2010}: 
\begin{equation}
S_{3D}=-\left(\frac{3}{2} +\lambda\right) \left(\frac{\pi}{3} \right) ^{\frac{2}{3}}\frac{k_B^2m_eT}{e \hbar ^2n_{eff}^{\frac{2}{3}}}
\end{equation}
The carrier density at high temperature ($n_{HT}$) and the loss of carriers due to the transition ($\Delta n_{eff}$) were estimated from the linear parts of $S(T)$ at high and low temperatures respectively, with an assumption, that the in - plane effective mass $m^*=m_e$ and is constant with T both for electrons and holes.
We have assumed the scattering factor $\lambda=-\frac{1}{2}$, corresponding to scattering with acoustic phonons. The obtained $n_{HT}$ - effective electronic density for $T>T^*$ is for each sample (see table \ref{seebecktable}) reasonably close to the values calculated from the chemical formula ($n_{chem}$). For $x$ = 1.16, one can compare the concentration of condensed carriers deduced from both the magnetic measurements and the Seebeck measurements. We find respectively $\Delta n= 1.12 \cdot 10^{27}m^{-3}$ and $\Delta n_{eff}=2.35 \cdot 10^{27}m^{-3}$, that is in rather good agreement. This shows, that the used model, despite of its roughness can serve as an useful guide in the estimation of carriers density in quasi 2D metals.
 \begin{table}[h]
 \caption{\label{seebecktable}Effective carrier concentrations in K$_{x}$P$_4$W$_{8}$O$_{32}$ for different doping, estimated from
$S(T)$ with semi - classical 3D model } 
 \begin{ruledtabular}\begin{tabular}{cccc} 

 K content & $n_{chem}$ [10$^{27}$ m$^{-3}$] & $n_{HT}$ [10$^{27}$ m$^{-3}$] & $\frac{\Delta n_{eff}}{n_{HT}}$[\%] \\ 
 \hline
0.93 & 7.88 & 2.72 & 22.20 \\
1.10 & 8.16 & 2.26 & 12.13 \\
1.16 & 8.28 & 2.35 & 34.16 \\
1.21 & 8.34 & 4.70 & 8.67 \\ 

\end{tabular}\end{ruledtabular}
 
 \end{table}
 The $\Delta n_{eff}$ found in compounds with visible hole contribution are notably higher than one found for $x$ = 1.21, where the electrons are dominant in whole temperature range. This effect can be explained as caused both by condensation of electrons, thus their removal from the conduction band and the increase of role of light holes, which contribute to n$_{eff}$ with sign opposite to electrons. 
 
 The investigation of the existence of CDW in low dimensional material must not omit the search for non - Ohmic DC transport. The evidence of Fr\"ohlich transport has already been reported in a number of quasi 1D materials as NbSe$_3$ \cite{monceau1976}, K$_{0.3}$MoO$_3$\cite{Dumas1983, Forro1986} or Tl$_{0.3}$MoO$_3$ \cite{collins1985}, while the attempts to observe CDW sliding in quasi 2D materials were unsuccessful even in powerful electric fields \cite{disalvo1980} with exception only to single compound, DyTe$_3$ \cite{sinchenko2012}. The difficulty with the study of this effect arises both from notably stronger pinning observed in 2D systems and from metallicity which remains in 2D materials after imperfect nesting of the Fermi surface, which obstructs reaching strong electric fields. We have performed the measurement of the $I(V)$ curve in 
K$_{1.15}$P$_4$W$_{8}$O$_{32}$ sample cooled in liquid nitrogen applying DC current up to $I$ = 6 A, reaching maximum electric field of $E$ = 121 mV/cm. No break of linearity in DC transport was found in this temperature range.  We also emphasize, that while the CDW nesting vector doubles the cell periodicity, the depinning energy can be significantly  increased by the commensurability term \cite{lee1979, Lee1974, Dumas1986} which can even prevail over the impurity pinning as in quasi 1D (2,5(OCH$_3$)$_2$ DCNQI)$_2$Li \cite{tomic1997}.
Note, that the maximum $E$ we have reached is a factor of 5 smaller than the minimum threshold field in DyTe$_3$ which rises further at temperatures far below $T^*$ and the commensurability term in  K$_{x}$P$_4$W$_{8}$O$_{32}$ is expected to raise the pinning potential to the values comparable to the electronic gap. We suggest, that the lack of observed non-linearity is caused by strong pinning arising both from commensurate nature of the modulation and from the 2D electronic character of the tested material.  Therefore, the lack of non-linearity in DC transport can not contradict the CDW scenario in K$_{x}$P$_4$W$_{8}$O$_{32}$. 
\subsection{Discussion}
The nature of the transitions observed in K$_{x}$P$_4$W$_{8}$O$_{32}$ was a subject of an intensive dispute. Drouard et. al \cite{Drouard1999}
 remarked, that the modulation vector should significantly vary with $x$, while for each $x$ one observes the identical $q$ = 0.50 a$^*$ modulation vector. On the contrary, the band structure calculations performed by Canadell \textit {et al}\cite{canadell1991}  predict, that the modulation vector associated with hidden nesting of quasi 1D part of the K$_{x}$P$_4$W$_{8}$O$_{32}$ Fermi surface varies from 0.47 a$^*$ to 0.50 a$^*$ for $x$ = 0.8 and $x$ = 3.5 respectively and this deviation is far from significant. Bondarenko {\textit {et al}\cite{bondarenko2004} concluded, that the large (in comparison to the undoped $m$ = 4) change in the specific heat due to transition observed at $T^*$ indicates that the transition is essentially structural and not a simple CDW transition.  Nevertheless, we find the comparison between $\Delta C_p/Rp$ ($p$ is the number of electrons per formula unit and 
$R$ = 8.31 J mol$^{-1}$ K$^{-1}$ is the gas constant) in K$_{x}$P$_4$W$_{8}$O$_{32}$ and corresponding quantities in 'well established' CDW materials misguiding. Considering both electrons donated by P groups and K atoms in the tunnels, the value of p should denote 5.07, 5.30, and 5.45 for $x$ = 1.07, $x$ = 1.30 and $x$ = 1.45 respectively, instead of values slightly larger than unity considered by Bondarenko \textit {et al}. Then, we find that  $\Delta C_p/Rp$ for K$_{x}$P$_4$W$_{8}$O$_{32}$ should vary between 0.27 and 0.47 for $x$ = 1.07 and $x$ = 1.45 respectively, which is finally smaller than the values found in examples of known CDW materials  K$_{0.3}$MoO$_3$ (0.53), KMo$_6$O$_{17}$ (0.40) or $\eta-$Mo$_4$O$_{11}$ (0.78), recalled by Bondarenko \textit {et al}. Then, the argument that $\Delta C_p/Rp$ is too large to be attributed to a single CDW transition appears not very strong. Dusek \textit {et al}\cite{Dusek2002} suggested, that the transition is driven by atomic displacements due to strain induced between K atoms and the PO$_4$ tetrahedra instead of the CDW formation. We agree, that this mechanism would produce a change in the electronic band structure, but its effect is not essential to explain the physical properties showing modification of the Fermi surface accompanied with carrier condensation.
The small hysteresis observed in resistivity is an indicative of a first - order transition, suggesting the strong lattice component of the observed anomaly. Such scenario is expected in a strong coupling approach \cite{McMillan1977}. On the other hand, this model predicts the electronic gap to be substantially larger than the weak coupling value of 2$\Delta=$ 3.52 $kT^*$ \cite{Kwok1990, Smontara1992}. This is not relevant in K$_{x}$P$_4$W$_{8}$O$_{32}$, since our results show the excellent agreement between the measured electronic gap and the weak coupling prediction. In the strong coupling scenario, at $T>T^*$, the short order fluctuations are still preserved leading to diffuse scattering in the vicinity of superlattice reflections observed below $T^*$ \cite{Rossnagel2011} which is not visible in our sample even with synchrotron radiation. Another interesting point is also the weakness of structural distortion in the W-O assembly (displacements in W-O and W-W distances are $\approx$ 0.01 to 0.02 \AA) associated with the CDW formation in MPTB\cite{Foury20022, roussel2001}, which does not corroborate with the requirement of large distortions in strong coupling mode. The BCS character of condensed electronic density and electronic gap are strong arguments for a genuine second order nature of the electronic part of the transition. Note, that the hysteresis is not visible in magnetic susceptibility. Then, it is reasonable to assume, that the hysteretic behavior of resistivity likely originates from the structural distortion in K-P-O environment which produces an additional effect on the scattering term. In this picture, if one considers the residual strain between K atoms and PO$_4$ tetrahedra as a main driving force of the structural transition in K$_{x}$P$_4$W$_{8}$O$_{32}$, the Peierls instability can be seen as a by - product of the structure modification. The evolution of a band structure upon the transition allows the nesting of a 1D fragment of the Fermi surface with a preferential wavevector $q=0.5a^*$, independent of $x$ and overcome the suppression of the CDW \cite{Pouget1989} due to the disorder caused by inhomogeneous potassium distribution. Then, it is favorable to maintain this commensurate value unchanged with K content to preserve the optimal nesting conditions. This sort of resonance of CDW and the underlying lattice leading to existence of one privileged modulation vector can partially explain the lack of subsequent nesting of remaining Fermi surface parts as predicted by Canadell and Whangbo \cite{canadell1991} or observed in undoped P$_4$W$_{8}$O$_{32}$ \cite{dumas2002}.

\section{conclusions}
In this article, we report a full set of experimental data on K$_{x}$P$_4$W$_{8}$O$_{32}$,
including new high resolution X-Ray data, magnetic susceptibility, 
and detailed chemical characterization to focus on the nature of
the transition observed at $T^*$. We conclude, that the anomalies observed in physical properties of  K$_{x}$(PO$_{2}$)$_{4}$(WO$_{3}$)$_{8}$ are associated with condensation of
around 15 \% of the total number of carriers and the hidden nesting of the quasi 1D portion of the Fermi Surface. We also find the enhancement of transport properties anisotropy upon the FS modification.  Despite the resistivity hysteresis, we have provided the strong evidence for the weak coupling Peierls scenario. We propose that the thermal hysteresis  does not arise from a first order characteristics, but from scattering due to residual strains around the structural transition, which also modifies the Fermi surface and enables the second order charge density wave instability.
\\

\begin{acknowledgments}
Financial support by the French National Research Agency
ODACE ANR-Project number 2011-BS04-004-01 is gratefully
acknowledged.
\end{acknowledgments}

%

\end{document}